\documentclass[prl,twocolumn,showpacs,preprintnumbers,amsmath,amssymb,superscriptaddress]{revtex4-1}
\usepackage{epsfig}
\usepackage{graphicx,amssymb,epsfig,amsmath}
\usepackage{color}
\usepackage[normalem]{ulem}

\def\tblue{}
\def\tgreen{}

\newcommand{\cred}[1]{{#1}}

\newcommand{\ICP}{{\it{}ICP}}
\newcommand{\LEHV}{{\it{}ERT}}
\newcommand{\GPT}{{\it{}GPT}}

\newcommand{\RAC}{{\it{}RAC}}
\newcommand{\be}{\begin{eqnarray}}
\newcommand{\ee}{\end{eqnarray}}

\newcommand{\removed}[1]{}

\begin{document}

\title{Information Content of  Systems as a  Physical Principle}

\author{L. Czekaj}
\affiliation{Institute of Theoretical Physics and Astrophysics, National Quantum Information Centre, Faculty of Mathematics, Physics and Informatics University of Gda\'nsk, 80-308 Gda\'nsk, Poland}

\author{M. Horodecki}

\affiliation{Institute of Theoretical Physics and Astrophysics, National Quantum Information Centre, Faculty of Mathematics, Physics and Informatics University of Gda\'nsk, 80-308 Gda\'nsk, Poland}

\author{P. Horodecki}

\affiliation{Faculty of Applied Physics and Mathematics, Gda\'nsk University of Technology, 80-233 Gda\'nsk, Poland}
\affiliation{Institute of Theoretical Physics and Astrophysics, National Quantum Information Centre, Faculty of Mathematics, Physics and Informatics University of Gda\'nsk, 80-308 Gda\'nsk, Poland}

\author{R. Horodecki}

\affiliation{Institute of Theoretical Physics and Astrophysics, National Quantum Information Centre, Faculty of Mathematics, Physics and Informatics University of Gda\'nsk, 80-308 Gda\'nsk, Poland}

%\author{K. Horodecki}
%\affiliation{Institute of Informatics, University of Gda\'nsk, 80-952 Gda\'nsk, Poland}
%\affiliation{National Quantum Information Centre in Gda\'nsk, 81-824 Sopot, Poland}

\date{\today}
\begin{abstract}
To explain conceptual gap between classical/quantum and other, hypothetical descriptions of world, several principles has been proposed.
So far, all these principles have not explicitly included the uncertainty relation. 
Here we introduce an information content principle (ICP) which
represents the new - constrained uncertainty principle. The principle, by taking into account
the encoding/decoding properties  of {\it single} physical system, is capable of  separation both classicality and quanta from a number of  potential  physical theories including hidden variable theories.
The ICP, which is satisfied by both classical and quantum theory,  states that the amount of non-redundant information which may be extracted from a given system is bounded by a perfectly decodable information content of the system. We show that ICP allows to discriminate theories which do not allow for correlations stronger than Tsirelson's bound.
We show also how to apply the principle to composite systems, ruling out some theories despite their elementary constituents behave quantumly.
\end{abstract}

\maketitle

{\it Introduction}. -
%problem description
%We understand mathematics of quantum mechanics quite well but we have problem with understanding quantum mechanics itself.
%The mathematics of quantum mechanics is understood quite well but there is a problem with understanding quantum mechanics itself.
It is astonishing that our best theory of the fundamental laws of physics, quantum mechanics being robust against innumerable experimental tests is as well robust against our understanding of its physical origins.
This is notoriously manifested by the variety of interpretations of quantum mechanics  (e.g. \cite{cabello_mam}). One of the reasons is the way the postulates of quantum mechanics are expressed: they refer to highly abstract mathematical terms without clear physical meaning. This drives physicists to look for an alternative way of telling quantum mechanics.The problem was attacked on different levels. On one hand it has been shown that quantum theory can be derived  from  more intuitive axioms
%There were attempts of deriving quantum theory from more intuitive axioms \cite{qm_derive_mackey}
%qm_deriv_vn,
\cite{Weizsacker,qm_derive_mackey,qm_without_amplitudes,Zeilinger,qm_deriv_hardy,Brukner_Zeilinger,Dakic_Brukner,qm_deriv_mm,qm_deriv_chiribella,qm_deriv_1}). In particular it is related to the vastly developed field aiming at reconstructing quantum theory from information properties of the system
%\cite{Weizsacker,Zeilinger,Brukner_Zeilinger,Dakic_Brukner,qm_deriv_1}
\cite{Zeilinger,qm_deriv_hardy,Brukner_Zeilinger,Dakic_Brukner,qm_deriv_mm,qm_deriv_chiribella,qm_deriv_1}
(cf. \cite{Griffiths1}).
%,qm_deriv_h2
 On the other hand an effort was made to derive some principles
\cite{inf_causality,macroscopic_locality,Cabello_exclusivity,local_orthogonality}
which can separate quantum theory (or in a narrow sense some aspects of the theory, such  as correlations) from so called \cred{\it super-quantum theories} i.e. the theories that inherit from quantum theory the no-signaling principle,
but otherwise can offer different predictions than quantum mechanics \cite{PR}.
%The main purpose of the present paper 

However, in difference to those approaches, our goal is to find  a criterion for physical  theories  %as an elementary information principle
which involves quantitative rather than qualitative (i.e.logical) constraints.
Furthermore, in contrast to the previous information principles based on composite system here we define and study a principle that refers to a {\it single} system and represents a sort of uncertainty principle. 

To this end we propose a constraint that ties together (i) the amount of {\it non-redundant information} which  can be extracted from the system by the set of observables and (ii) systems' informational content understood in terms of maximal number of bits that may be encoded in the system in perfectly decodable way. We call this constraint {\it information content principle} (\ICP{}) which
represents the new - {\it constrained} uncertainty principle which \tblue{holds in} the classical and quantum theories for two different reasons. For classical systems it is due to lack of knowledge while for quantum systems it reflects \tgreen{quantum uncertainty} \cite{Maassen-Uffink}.
To demonstrate the efficiency of the principle we show how it is violated by some theories with relaxed uncertainty constraints \cite{SW_relaxed_uncertainty}  and polygon theories \cite{polytopic_models} as well as some {\it incomplete} classical theories \tblue{ akin to epistemically restricted theories \cite{hbit_toy_model,spekkens_epist_rest}.}

{\it Information content principle}. -
Our aim is to provide a principle that would bound the information extractable from a system in a physical theory.
There are two problems here to address: firstly, what should be the ultimate bound for such extractable information,
and second, how the extractable information itself is to be defined.

Regarding the bound, it is natural to employ the following fundamental
quantity, which we call {\it information content}.
The information content is {\it the maximal number of bits, that can be encoded in a lossless way into a given system}. \cred{ We express it as $\log_2 d$ where $d$ the maximal number of messages that can be encoded in a lossless way into a system 
(see Appendix \ref{app:generalized_prob_theories}  for detailed discussion)}.
This is a quantity intrinsic to
any given theory, for example in quantum mechanics, it is given by the logarithm of the dimension of the Hilbert space of the system.

The second question is more demanding. We shall first present some rough picture, and then propose a concrete
implementation of the idea. To begin with, information can be extracted from the system by making measurements.
Rather than trying to determine full amount of extractable information, we will consider information obtained from
measuring some set of observables. We might want to add informations extracted by measuring each observable, however they may be redundant
(e.g. if one observable is a function of other observable).
Therefore, one has to subtract the redundancy. This can be symbolized by the following expression:
\be
\label{eq:main}
\sum_i I_{M_i} - I_R \leq I_C
\ee
where
$I_{M_i}$ denotes information obtained by measuring observable $M_i$, and $I_R$ represents redundant information
and $I_C$ is the total information content as defined above. Now we would like to make the above formula more concrete, so that all the quantities can be computed in a given theory.

To see the difficulties which arise, when one tries to define redundant information
$I_R$, consider two observables $M_1$, $M_2$ in classical theory. Then a natural candidate is just the mutual information 
of the joint probability of the outcomes of $M_1$ and $M_2$
\be 
I(M_1:M_2)=H(M_1)+H(M_2) - H(M_1,M_2)
\ee
(here $H(\{p_i\})=-\sum_ip_i \log_2 p_i$ denotes Shannon entropy of probability distribution $\{p_i\}$; thus
$H(M_1)$ and $H(M_2)$ are entropies of distributions of $M_1$ and $M_2$, respectively, while $H(M_1,M_2)$ is 
entropy of joint probability distribution of $M_1$ and $M_2$). Indeed, the mutual information can be interpreted as a common information
shared be both random variables. However, in quantum theory, such joint probability does not exist.
Therefore, since our quantities are to be sensible in any theory, including the quantum one,
we have to define redundancy in some indirect way.

\begin{figure}[htbp]
	\centering
		\includegraphics[scale=0.4]{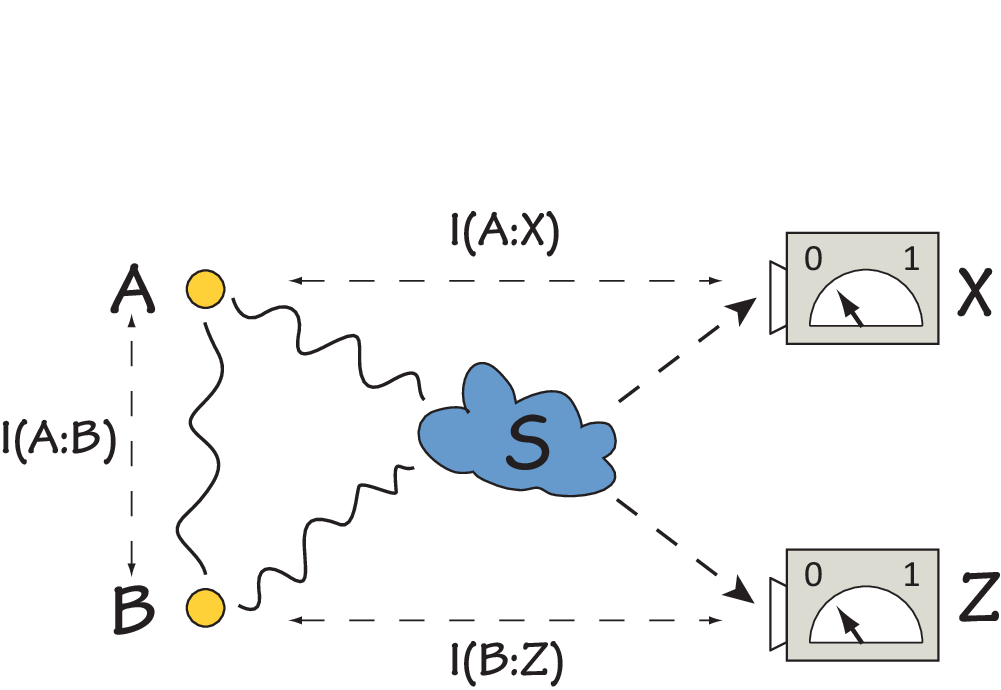}
	\caption[\ICP scenario]{\label{fig_2_icp_scenario}\textbf{Scenario of the Information Content Principle} }
\end{figure}

We shall now present a setup which allows to properly grasp  the idea of non-redundant extracted information.
Consider the following scenario (for simplicity, just for two measurements) depicted in Fig.
\ref{fig_2_icp_scenario}.

Consider two persons: the sender and the receiver. 
Let the sender  holds classical information stored in two registers $A$ and $B$. She wants to provide access to that information for the sender but she does not know which register is interesting for him.
	She prepares the system $\mathcal{S}$ in state which depends on the content of $A$ and $B$. Then she sends $\mathcal{S}$ to the receiver. 
After transmission of $\mathcal{S}$, the sender and the receiver share the state 
\be
\label{eq:state}
\omega^{SAB}=\sum_{i,j}p_{i,j}\omega^{S}_{i,j}\otimes\sigma^A_i\otimes\sigma^B_j
\ee
 where $p_{i,j}$ is distribution of the content of the classical registers, whose states are here labeled as $\sigma^A_i$ and $\sigma^B_j$;
$\omega^S_{i,j}$ denotes the state of the system, given the classical registers are in the state $\sigma^A_i\otimes \sigma^B_j$;
the classical registers $A$ and $B$ are in hands of the sender, while the system $S$ is held by  the receiver. 
	The receiver extracts information from $\mathcal{S}$ by performing one of two %incompatible
measurements $X$ and $Z$. In this way he learns about the content of $A$ or $B$ respectively.
The information extracted by observables $X$ and $Z$ are defined respectively as $I(A:X)$ and $I(B:Z)$ - the Shannon mutual information
between classical system and outcomes of measurement. The redundant information will be the mutual information
between the classical systems $I(A:B)$. The formula \eqref{eq:main} then takes the following concrete form
\cite{KH_unpublished} (see also
\cite{KH_RACBOX}).
\begin{equation}
I(X:A)+I(Z:B) - I(A:B) \leq \log_2 d\label{eq:principle_conc},
\end{equation}
where $\log_2 d$ is the \cred{information content of a system}.
Note, that here all the quantities $I$ are mutual informations of classical variables.
For more than two measurements $\{X_i\}$ the formula takes the form (see Appendix \ref{supplement material_ICP_proof} 
 for details):
\begin{equation}
\sum_i I(X_i:A_i)-I(A_1:\ldots:A_n) \leq \log_2 d\label{eq:principle_conc_multiple}.
\end{equation}

Now, the central  postulate of the present paper is that the above formula
\eqref{eq:principle_conc_multiple} represents  the information content principle which should be valid for {\it  any}
physical system (either an elementary or a composite one) in {\it any}  physical theory.
In particular, the principle holds for quantum theory, which can be proved in spirit of \cite{inf_causality}. Namely,
any theory in which one can define a notion of entropy satisfying some natural axioms obeys the principle. In Appendix \ref{supplement material_ICP_proof}  we give the list of axioms, and derive ICP from those axioms.
The axioms are satisfied by von Neumann entropy in quantum theory and by Shannon entropy in classical theory,
hence both theories obey the principle.

Note that the \ICP{} incorporates idea of impossibility of encoding more information using complementary observables \tgreen{\cite{RAC_intro}}
 which  is a basic ingredient of information-type principles
	\cite{Dakic_Brukner,qm_deriv_mm,Beigi,inf_causality,Pawlowski-Scarani-IC,Brunner_dimensions}
	(this idea differs from bounding capacity of quantum systems as well as bounding the classical memory
	required for their simulation, see. e.g. \cite{Cabello_memory}). 
	%\tgreen{More specifically we show that two thus far  distinct fundamental concepts of uncertainty and RAC are inextricably and quantitatively linked within ICP.}

Below we shall show violation of \eqref{eq:principle_conc} in two elementary examples:
(i) non-local theories represented here by so-called sbit (square-bit)~\cite{SW_relaxed_uncertainty};
(ii) epistemically restricted theories where as example we consider hbit (hidden-bit)~\cite{hbit_toy_model,spekkens_epist_rest}
and postpone discussion of more advanced cases to the further part of the paper.
To show violation for sbit and hbit, we evaluate \eqref{eq:principle_conc} on the state of \eqref{eq:state}
%\begin{equation}
%\omega^{SAB}=\sum_{i=0,j=0}^1 \frac{1}{4}\omega^{S}_{i,j}\otimes\sigma^A_i\otimes\sigma^B_j,
%\end{equation}
with $\omega^{S}_{i,j}$ being the such state of sbit or hbit 
that the outcome $i,j$ after measuring $X,Z$ respectively is certain (i.e. $p(a=i|x=X)=1$ and $p(a=j|x=Z)=1$).
Information encoded in the observables is completely uncorrelated i.e $I(A:B)=0$ for $\omega^{SAB}$. Since there is no uncertainty in the system, information encoded in each observable might be recovered completely hence $I(X:A)=I(Z:B)=1$. Taking that together we obtain violation of \eqref{eq:principle_conc} since $I(X:A)+I(Z:B)-I(A:B) = 2 > 1$. At this point it is worth to notice that in the case of hbit, violation comes from the fact, that the observed dimension $d$ is different from what we could call
 "intrinsic" system dimension: the observables available in theory have two outputs while the internal state of the system is 
described by $2$ classical bits. The theory is incomplete because of lack of fine grained observable with four outputs that
could access full information available in the system. Note that, lack of such observable excludes possibility of measuring the two observables one after another (in which case, one would eventually access both bits): indeed, then one could define the fine grained observable 
as the subsequent measurement of the dichotomic observables.

For classical bit, if the observables are nontrivial, they must be a function of one another - so that we have actually only one observable up to relabeling the outputs, and the information is highly redundant. Indeed, if $I(A:X)=1$ and $I(A:Z)=1$, then we must have $I(A:B)=1$.
Both informations are maximal, but they are redundant.
To discuss  the quantum case, let us assume that marginal entropies $H(A)=H(B)=1$. Then for $X$ and $Z$ complementary, we have that $I(A:B)$
can vanish. Hence we have $I(A:X)+I(B:Z) - 0 \leq 1$. Thus, although the informations are non-redundant, they are restricted.
Thus unlike in classical case, here we have two independent ``species'' of information and there is room only for one of them.
If we rotate observable $Z$ towards $X$, we observe that $I(X:A)+I(Z:B)$ grows up together with $I(A:B)$. The observables disclose more information, however the information is more redundant. Extractable information cannot exceed the bound given by \ICP{}.

To see that \ICP{} can be interpreted as an uncertainty of a new kind,
suppose that we fix $I(A:B)$ to be some number strictly less than 1, i.e.
\be
I(A:B)<1.
\ee
%{\color{blue}
Then we obtain restrictions on the values $I(A:X)$ and $I(B:Z)$, namely
they cannot be both equal to $1$. This would look like Hall's exclusion principle
which also bounds the sum  of two mutual informations \cite{IEP}.  However, unlike in exclusion principle, in the present case the restriction is the same regardless  the observables commute or not.
Indeed, if the observables commute (i.e. in classical theory) the restriction comes from the fact, that up to the relabelling of outcomes, there is only one fine-grained observable on a classical system, so that any other observable carries the same information.
Therefore it is impossible to fit more than one bit into a binary system, as \ICP{} states.

If the observables do not commute, the reason is less  obvious, because the different observables, especially if they are complementary,
surely do not carry the same information. However, again the restriction posed by \ICP{} holds, this time because of
the \tgreen {quantum} uncertainty.
%}

To see more clearly the connection with quantum uncertainty, let us assume that $H(X)=1$ and $H(Z)=1$ i.e. that the outcomes are random.
Then \ICP{} writes as
\be
H(X|A) + H(Z|B)  \geq 1 - I(A:B).
\ee
While the standard \tgreen{quantum} uncertainty principle is of the form \cite{Maassen-Uffink}
\be
H(X|A) + H(Z|B)  \geq c(X,Z)
\ee
where $c(X,Z)$ quantifies the lack of common eigenvector for $X$ and $Z$. For commuting observables $c=0$, and the uncertainty relation
is trivial - i.e. there no uncertainty. In our case, when $I(A:B)<1$, the right hand side is constant independent of observables,
hence the relation is {\it always - both classically and quantumly} -- nontrivial. Thus any theory which obeys \ICP{}, exhibits uncertainty of outcomes, under
the constraint $I(A:B)<1$. Yet, as we show below, there are theories that do not exhibit this uncertainty, and in this sense they are too ``certain'' to be physical.

\begin{figure}[htbp]
	\centering
		\includegraphics[scale=0.30]{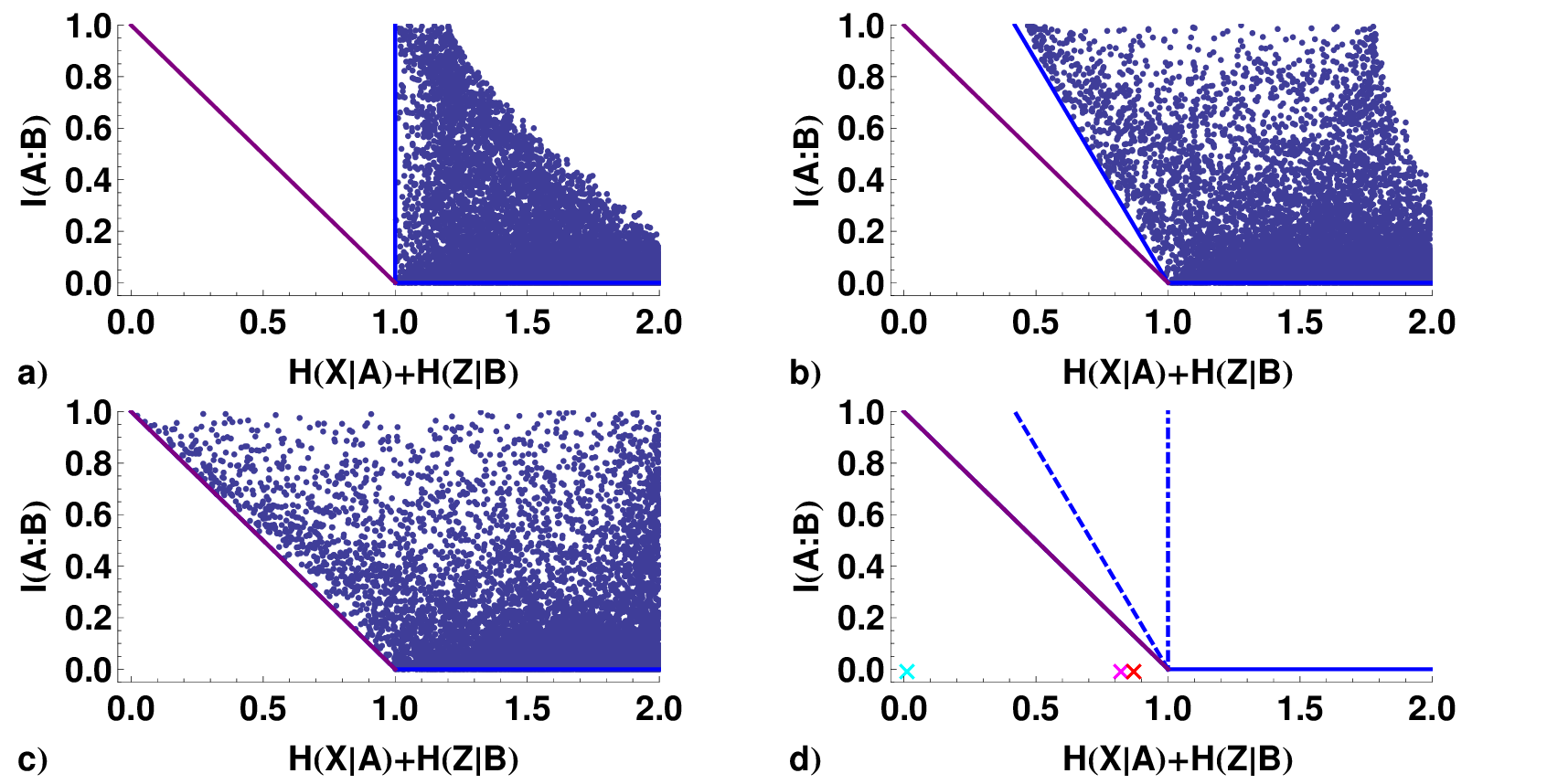}
	\caption[\ICP as constrained uncertainty principle]{\label{fig_3_cutHXAHXB}
	\textbf{ICP as constrained uncertainty principle}
In the picture we presents on the plain
$H(X|A)+H(Z|B) \sim I(A:B)$ the areas attainable by quantum 1-qbit system for some fixed observables $X$ and $Z$: (a) complementary observables; (b)  observables with the angle between axes representing them  on the Bloch sphere is $\pi/4$ (they are still noncommuting,
but are not complementary anymore);
(c) commuting observable (X=Z), which we interpret as classical case. Blue dots corresponds to states $\rho_{SAB}$ chosen randomly from set of states satisfying $H(X)=H(Z)=1$. The blue solid line represents schematically boundary of the area. The purple line depicts ICP bound. The area attainable by quantum and classical theory is placed above this line. In item (d) we put together the areas from (a), (b) and (c). We can observe that in setup when registers $A$ and $B$ keep completely independent information (i.e. $I(A:B)=0$) in both, classical and quantum case there is unavoidable uncertainty. However when registers $A$ and $B$ hold the same information, uncertainty in classical case vanishes. On the other hand, in some non-local theories like polygon theories, for $I(A:B)=0$ there are states that are "more certain" than classical and quantum ones. These states were depicted by "x" on the item (d) and correspond to polygon theories with parameter $n=4$ (cyan), $n=6$ (magenta), $n=8$ (red).
	}
	
\end{figure}

The above considerations are illustrated in the Fig.~\ref{fig_3_cutHXAHXB}.

%}

{\it Violations of} ICP {\it in general probabilistic theories}. -
{%
In this section we briefly discuss violation of \ICP{} in two families of theories that originate from \GPT{}. 
We start with {\it p general non signaling theories} (p-GNSTs) introduced in \cite{SW_relaxed_uncertainty} (for more details see Appendix \ref{app:generalized_prob_theories}}).
These theories violate the quantum uncertainty relation for anti-commuting
observables %\cite{anticomm_ur}
and they were originally developed to study how
Tsirelson's bound for the CHSH inequality emerges from the uncertainty relation.

Elementary system of p-GNST theory is a box with two observables $X$ and $Z$. Its state space is bounded by uncertainty relation:
\begin{equation}
(s_x)^p+(s_z)^p \leq 1,
\end{equation}
where $p\in[2,\infty]$ is parameter of the theory and $s_x=p(a = +|x = X)_\psi-p(a = -|x = X)_\psi$ is the mean value of observable $X$ measured on the system in state $\psi$ (analogically for $s_z$ and $Z$). Varying parameter $p$ one can move from state space of sbit to qbit.

In Appendix \ref{sec:pGNSTdetails} we show that violation of quantum uncertainty relation by states from p-GNST 
(i.e. for theories with $p>2$) not only leads to violation of Tsirelson's bound (as proved in \cite{SW_relaxed_uncertainty}) but also to violation of \ICP{} by elementary system.
Violation of \ICP{} in these theories follows from existence of sufficiently many  states for which  entropic uncertainty for observables $X$ and $Z$ is smaller than in quantum case.
Relaxation of uncertainty relation also was shown to increase of maximum recovery probability for 
so called {\it $2\mapsto 1$ random access codes} (RAC). Namely, one can encode two bits into a system from the theory, 
in such a way, that  probability of decoding (recovery) for each bit separately 
reads as $p_{rec}=\left(1/2\right)^{1/p}$. Therefore, excluding p-GNSTs with $p>2$, \ICP{} puts bound on performance of random
access codes.

\ICP{} not only applies to elementary systems but may be also used in natural way
to study composite system. It is able to exclude theories which are non-physical nevertheless their state space of elementary system is quantum.
Here an example is p-GNST with $p=2$ where violation of \ICP{} occurs for system with at least 5-parties. This result bases on the existence of super strong \RAC{} in p-GNST.

It is interesting to ask what other geometrical constraints (e.g. other uncertainty relations, consistency constraints (cf.~\cite{SW_relaxed_uncertainty}), local orthogonality \cite{local_orthogonality}) have to be added to theory to conform with~\ICP{}. In the opposite direction, one may ask how \ICP{} limits strength of non-local correlations achievable in \GPT{} for system whose elementary subsystems obey quantum mechanics.

%In this way we shed some light on connection between these concepts.
%On the ground of these theories, we can also discuss bound on  maximum recovery probability of quantum $2\mapsto 1$ \RAC{}.
%{ powiedziec cos o streeringu w tych teoriach i remote measurement preparation??}

We move to polygon theories \cite{polytopic_models}. They are described in more detail in Appendix E2. Here we just mention that state space in that theories is given by polygon with $n$ vertices.
% and show that \ICP{} is also violated in the non-classical cases.
For those theories \ICP{} is {\it more sensitive} than Tsirelson's bound since it allows to discriminate theories which do not allow for correlations stronger than Tsirelson's bound.
Namely, \ICP{} is violated in all non-physical polygon theories.
For theories with even $n$, it is again connected with existence of states with lower entropic uncertainty than in quantum case. For odd $n$ violation of \ICP{} links rather to the fact that in this case polygon theories allows for communication of more than $1$-bit per elementary system in Holevo sense (i.e. in asymptotic limit) \cite{GPT_holevo_polygon}. Interestingly, we found examples of polygon theories that
are not ruled out by principle proposed in an independent development
\cite{Brunner_dimensions} based on so called {\it dimension mismatch}.

%%%%%%%%%%%%%%%%%%%%%%%%%%%%%%%%%%%%%%%%%%%%%%%%%%%%%%%%%%%%%%%%%%%%%%%%%%%
% summary
%%%%%%%%%%%%%%%%%%%%%%%%%%%%%%%%%%%%%%%%%%%%%%%%%%%%%%%%%%%%%%%%%%%%%%%%%%%
{\it Summary}. -
We have identified constrained uncertainty informational principle (ICP) based on the single physical system which puts new constraints for physical theories. 
%the physical principle (ICP) that involves an information content of the physical system asa single entity. ICP has operational character;it can be expressed in terms of Shannon mutual informations and classical outputs of measurements. It  can be alsoapplied to composite systems.
The principle has a form of the uncertainty-type inequality with an extra  information constraint.
%The latter feature ie. the deliberate presence of constrains
%decides - up to our knowledge -  about novel character of it.
This is the feature that allows the principle to
filter out both \tblue{and ,,super-quantum''and ,,super-classical''(epistemically - restricted)} theories leaving the two ,,modest'' ones i.e. classical and quantum within the scope of its validity. If applied to classical theory \ICP{} reflects the fact that there is basically one type of information and all fine grained observables in classical discrete system are equivalent up to relabeling. On the contrary, in quantum mechanics, there are much different "species" of information which is reflected by the presence of incompatible observables which are only partially redundant.
At the same time only one type of information may be completely present in the system \cite{incomp_inf_1,incomp_inf_2} as it is stated in Bohr's principle of complementarity\cite{bohr_uncert}
which is connected to entropic uncertainty relations.
The only one observable from the complementary set may be measured perfectly.
However, two observables which are "less incompatible" than complementary, reveal information which is redundant.
\ICP{} gives the trade-off between how much of  information may be extracted and how redundant the information is. In particular, the power of the principle is illustrated by the fact that, it rules out some theories \tblue{(so called polygonic theories)} that do not violate Tsirelson bound, which therefore are not detected neither by information causality nor by local orthogonality.
Intriguing novel feature of the {\it ICP} is that two thus far distinct fundamental concepts uncertainty and random access coding are inextricably and quantitatively linked within {\it single} constrained uncertainty. It sheds light on the question why quantum mechanics is so restrictive or in other words: why it has such and only such strength.  
%In fact, it seems to be quite surprising, yet -  as we see here - still possible that the two theories, so different in their physical nature,  satisfy the principle that has the concept of uncertainty in its core.
%elementary system information content principle,
%that involves quantitative - uncertainty type constraints.
%It is used as a kind of the Ockham razor, which rules out unphysical theories.
%\mh{ten razor to czysta retoryka...}
%%\removed{The principle  is respected  by classical and quantum theories
%%and is violated by incomplete classical theories as well as postquantum ones: p-GNST and polygon theories.
%%The heart of the principle is uncertainty relation. \ICP{} can rule out some theories that do not
%%violate Tsirelson bound, which therefore are not detected by information causality and local orthogonality.}
%It is applicable to elementary as well composed systems.
%\tred{
%Moreover - to our knowledge - \ICP{} is the first principle that is capable to exclude epistemically-restricted theories admitting the existence of essentially hidden variables \tblue{(i.e. the variables, that cannot be physically accessed)}.
We believe that the information content principle may be an useful tool for analysis of the forthcoming theories or yet to be discovered. It seems also, that it may help in deeper understanding of the laws governing
physical reality in general.

The authors  thank A. Grudka, K. Horodecki, M. Pawlowski and R. Ramanathan
 for many discussions about information principles. 
 We also thank R. Augusiak, C. Brukner, N. Brunner, A. Cabello,
 R. B. Griffiths and L. Masanes for valuable comments and feedback
 on first version of the manuscript.
 This paper is supported by Polish Ministry of Science and Higher Education
Grant no. IdP2011 000361,  ERC QOLAPS and Polish National Science
Centre  Grant  no. DEC-2011/02/A/ST2/00305, and partially by the John Templeton Foundation
the grant ID 56033.

%\begin{supplemental material}
\begin{appendix}

\section{Generalized Probabilistic Theories\label{app:generalized_prob_theories}}
A generalized probabilistic  theory consists of a convex state space $\Omega\subseteq\mathbb{R}^n$ i.e. the set of admissible states the system may be prepared in, and the set of measurements $\mathcal{M}$. Measurement outcome is represented by effect $e$ which is linear map $e:\Omega\rightarrow[0,1]$. $e(\omega)$ is the probability of outcome $e$ when the measurement is performed on the system in state $\omega$. The special effect is the unit effect $u$ such that for every $\omega\in\Omega$ there is $u(\omega)=1$ (here we consider only normalized states).
The measurement is the set of effects $\{e_i\}$ summing up to unit effect $u$.
		
	%{We distinguished pure states and pure effects which are extremal points of the proper sets. Measurement which contains only pure effects is called pure measurement \cite{no_info_gain}.}
		
	The state of the system is entirely determined by the probabilities $p(a|x)$ it assigns to the outcomes $a$ of every measurement $x$. However there exist subset of measurements called fiducial measurements $\mathcal{F}\subseteq\mathcal{M}$ which is enough to describe the state\cite{qm_deriv_hardy}.
	
	%States which belongs to theory with $n$ fiducial measurements \cite{qm_deriv_hardy}.	
	%Since $\Omega$ is convex set, any statistical mixture of states is a state.
	
\begin{figure}[htbp]
	\centering
		\includegraphics[scale=0.4]{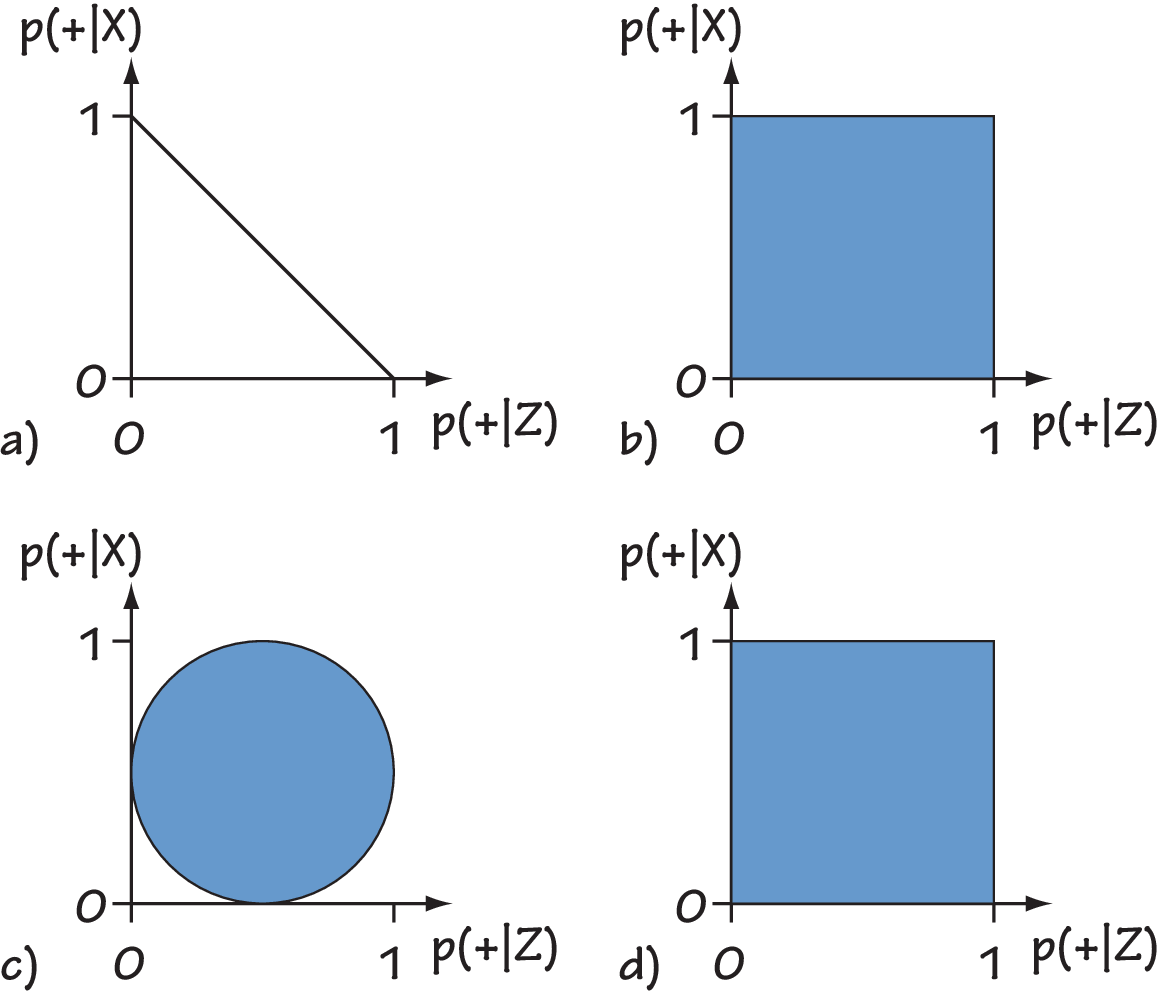}
	\caption[\GPT theories]{\textbf{Elementary systems in exemplary \GPT{} from the perspective of two distinguished dichotomic observables $X$ and $Z$.}  (a) classical bit, (b) hidden-bit (hbit): this system is an example from epistemically restricted theories, it consisting of two classical bits where if one of them, chosen by the observer, is read out then the second necessarily disappears or becomes unreadable by any physical interaction, (c) quantum bit (qbit) and (d)
square-bit (sbit): this is an example from non-local theories and may be viewed as a building block of PR-boxes. Qbit differs from the sbit and hbit by the amount of uncertainty. Classical bit admits no uncertainty, however $X$ and $Z$ revel {\it the same} information. This is reflected in perfect correlations of $X$ and $Z$. State space of sbit and hbit observed from perspective of two observables may seem to be equivalent. What is missing on that picture is the issue of decomposition into pure states.
Measurements $X$ and $Z$ has two outcomes $+$ and $-$. Axis represents probability of outcome $+$ when measurement $X$ or $Z$ is performed on system in given state.
%It is closer connected to remote task and in this way to Information causality \cite{inf_causality} as well as Tsirelson's bound %and steering of maximally ceratin states\cite{uncert_nonloc}.
\label{fig_1_gpt_theories_app}}	
\end{figure}

	Particular examples of systems which may be expressed in terms of \GPT{} (see Fig.~\ref{fig_1_gpt_theories_app}) are: classical {\it bit}, {\it qbit} and  {\it sbit} (square-bit). The last one, sometimes called gbit for generalized bit, is the building block of PR-box \cite{GPT_intro_3}.
\begin{figure}[htbp]
	\centering
		\includegraphics[scale=0.4]{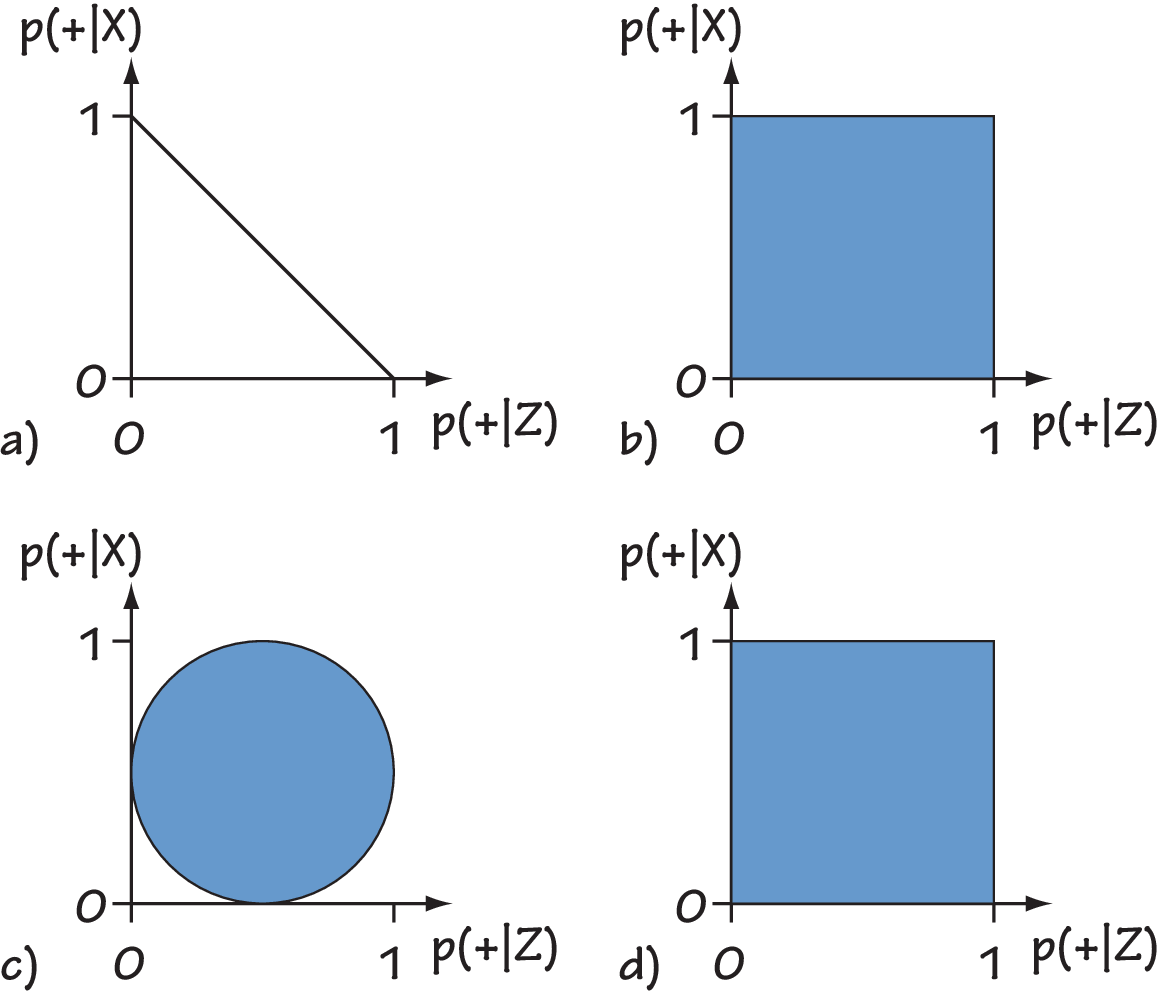}
	\caption[\GPT theories]{\textbf{Elementary systems in exemplary \GPT{} from the perspective of two distinguished dichotomic observables $X$ and $Z$.}  (a) classical bit, (b) hidden-bit (hbit): this system is an example from epistemically restricted theories, it consisting of two classical bits where if one of them, chosen by the observer, is read out then the second necessarily disappears or becomes unreadable by any physical interaction, (c) quantum bit (qbit) and (d)
square-bit (sbit): this is an example from non-local theories and may be viewed as a building block of PR-boxes. Qbit differs from the sbit and hbit by the amount of uncertainty. Classical bit admits no uncertainty, however $X$ and $Z$ revel {\it the same} information. This is reflected in perfect correlations of $X$ and $Z$. State space of sbit and hbit observed from perspective of two observables may seem to be equivalent. What is missing on that picture is the issue of decomposition into pure states. Measurements $X$ and $Z$ has two outcomes $+$ and $-$. Axis represents probability of outcome $+$ when measurement $X$ or $Z$ is performed on system in given state.
%It is closer connected to remote task and in this way to Information causality \cite{inf_causality} as well as Tsirelson's bound %and steering of maximally ceratin states\cite{uncert_nonloc}.
\label{fig_1_gpt_theories}}	
\end{figure}
	
If the set of measurements is not reach enough, one may obtain a classical system with hidden variables. For example,
elementary system in \LEHV{}  consists of two classical bits. One of two observables can be measured on system giving access to a chosen
bit. After measurement, information from the complementary bit is unavoidably  lost. This property reflects lack of fine grained observable in 
hidden variable theory. More sophisticated example of the hidden variable theory may be found in \cite{hbit_toy_model}.

Given two systems $A$ and $B$ we may define in \GPT{} a composite system $AB$.
The global state of the system $AB$ is completely determined by joint probabilities of outcomes for fiducial measurements performed at the 
same time on each subsystem. This is called local tomography assumption.
All effects for the composite system $AB$ are of the form $e_Ae_B$ which means that effect $e_A$ was measured on the subsystem system $A$ and $e_B$ on the subsystem $B$.
State space $\Omega_{AB}$ is not defined in the unique way. It contains all states of the form $\omega_A\omega_B$, i.e. states which  result from preparation of states $\omega_A$ and $\omega_B$ independently on the subsystems $A$ and $B$ (for $\omega_A\omega_B$ it holds $e_Ae_B(\omega_A\omega_B)=e_A(\omega_A)e_B(\omega_B)$). Other states $\omega_{AB}$ may also belong to $\Omega_{AB}$ provided $e_Ae_B(\omega_{AB})\geq0$ is true for every pair of effects.
Therefore starting from elementary systems, we may obtain different composite systems depending on the restrictions imposed on $\Omega_{AB}$ (cf. Generalized Non-Signaling Theory and Generalized Local Theory in \cite{GPT_intro_3}). In every case, $\Omega_{AB}$ contains only non-signaling states.
Dimension of state space $\Omega_{AB}$ is bounded by
\begin{equation}
\dim(\Omega_{AB})+1\leq (\dim(\Omega_A)+1)(\dim(\Omega_B)+1).\label{eq:state_space_dimm}
\end{equation}

%struktora efektow
%struktora stanow
%swoboda

%{\color{red} opisac bardziej spujnie observed system dimension i entropie}
For given \GPT{}, we may ask for maximal number of states that can be perfectly distinguished in a single-shot measurement \cite{qm_without_amplitudes,qm_deriv_hardy}.
We will call this value {\it observed system dimension} and denote it by $d$.
In terms of \GPT{}, we look for the biggest set $\{\omega_i\}\in\Omega$ such that there exist set of effects $\{e_j\}$ which obey $e_j(\omega_i)=\delta_{i,j}$. The set of states $\{\omega_i\}$ together with set of effects $\{e_j\}$ may be interpreted as a maximal classical subsystem of \GPT{} { and \cred{$\{e_j\}$ represents a generalisation of the quantum projective measurement, cf. complete measurement \cite{Distinguishability_measures}}}.
Observed system dimension is bounded by
\begin{equation}
d\leq \dim(\Omega)+1 \label{eq:dimbound}
\end{equation}
where equality holds only for classical systems\cite{GPT-fine_obs_MullerDV}. Combining~\eqref{eq:state_space_dimm} and~\eqref{eq:dimbound} one may obtain bound for composite systems.

%{
%Observed system dimension obeys multiplicativity rule, i.e. for composite system $AB$ we have $d_{AB}=d_{A}d_{B}$ where $d_A$ and $d_B$ observed dimensions of subsytems $A$ and $B$ respectively\cite{qm_deriv_hardy}.}

%composite system => multiplicativity d_{AB}=d_A d_B

%In hidden variables theories, some measurements are forbidden (they cannot be performed) however they are present in mathematical structure of theory. This leads to
%discrepancy between the number of states that can be perfectly disinguished with allowed measurements (i.e. observed dimension) and the number which is obtained if only mathematical structure of the theory is considered.

%%%%%%%%%%%%%%%%%%%%%%%%%%%%%%%%%%%%%%%%%%%%%%%%%%%%%%%%%%%%%%%%%%%%%%%%%%%
% ICP dla 2 observabli
%%%%%%%%%%%%%%%%%%%%%%%%%%%%%%%%%%%%%%%%%%%%%%%%%%%%%%%%%%%%%%%%%%%%%%%%%%%

\section{Information Content Principle for two observables\label{supplement material_ICP_proof}}

Here we provide detailed proof of \ICP{}. To make our argumentation easier to follow,
first we we consider only the case with two observables $\mathcal{M}=\{X,Z\}$. Then we generalize results to multiple observables scenario.

%Here we consider, for simplicity, only the case with two observables $\mathcal{M}=\{X,Z\}$. The scenario is depicted in \FIG{fig_2_icp_scenario}.
%For more general settings see SEC.~\ref{inf_principle_general}.

We start with definition of tripartite  state of the form:
\begin{equation}
\rho^{SAB}=\sum_{i,j}p_{i,j}\rho^{S}_{i,j}\otimes\sigma^A_i\otimes\sigma^B_j,
\label{eq:rghSAB}
\end{equation}
where state $\rho^S$ defined on system $\mathcal{S}$ belonging to considered theory (e.g. bit, qbit, sbit),
while $\sigma^A$ and $\sigma^B$ are classical registers. Their role is to keep classical information measured by observables $X$ and $Z$ respectively. The $\{p_{i,j}\}$ is classical probability distribution. The state $\rho^{SAB}$ is an analogue of quantum-classical system utilized in analysis of communication tasks.

%any $\rho^{SAB}$ where $\mathcal{S}$ is
We are in position to prove that~\eqref{eq:principle_conc_app} holds for  classical and quantum systems $\rho^{S}$:
\begin{equation}
I(X:A)+I(Z:B) - I(A:B) = I_C \leq \log_2 d\label{eq:principle_conc_app},
\end{equation}
where $I(X:A), I(Z:B), I(A:B)$ are classical mutual information %($I(A:B)=H(A)-H(A|B)$)
and $d$ is an observed system dimension.

%{\color{red}
Here we define mutual information and conditional entropy in the standard way as: $I(A:B)=H(A)-H(A|B)$ and $H(A|B) = H(AB) - H(B)$.
%defined as $\Iabs(S:F)=\Habs(S)-\Habs(S|F)$;
In the proof we make use from the following properties of classical and quantum entropies:
(i)  entropy of the system is bounded by $H(S)\leq\log_2 d$;
(ii) conditional entropy of any system $S$ correlated with classical one $C$ is non-negative
$H(S|C)\geq0$
% $\Habs(S|C)\geq0$
where $C$ is classical system;
(iii) strong subadditivity
%$\Habs(SA)+\Habs(SB)\leq\Habs(SAB)+\Habs(S)$
$H(SAB)+H(S)\leq H(SA)+H(SB)$;
(iv) information processing inequality for measurement
%$\Iabs(S:A)\geq\Iabs(X:A)$
$I(S:A)\geq I(X:A)$
where $X$ denotes measurement outcome.
In Appendix ~\ref{sec:entropy_props} we discuss these properties on the ground of \GPT{}.
%}

First, we use (i) and (ii) to obtain upper bound for mutual information between system $\mathcal{S}$ and $\mathcal{AB}$ for a state $\rho^{SAB}$:
%\begin{eqnarray}
%\Iabs(S:AB)&=&\Habs(S)-\Habs(S|AB)\nonumber\\
%&\leq&\Habs(S)\nonumber\\
%&\leq&\log_2 d\label{icp_up}.
%\end{eqnarray}

\begin{eqnarray}
I(S:AB)&=&H(S)-H(S|AB)\nonumber\\
&\leq&H(S)\nonumber\\
&\leq&\log_2 d\label{icp_up}.
\end{eqnarray}

Using chain rule for mutual information, we get
%\begin{eqnarray}
%\Iabs(S:AB)&=&\Iabs(S:A)+\Iabs(S:B|A)\nonumber\\
%\Iabs(AS:B)&=&\Iabs(S:B|A)+I(A:B).
%\end{eqnarray}
\begin{eqnarray}
I(S:AB)&=&I(S:A)+I(S:B|A)\nonumber\\
I(AS:B)&=&I(S:B|A)+I(A:B).
\end{eqnarray}

Putting this together with strong subadditivity and information processing inequality for
 measurements, we obtain
%\begin{eqnarray}
%\Iabs(S:AB)&=&\Iabs(S:A)+\Iabs(AS:B)-I(A:B)\nonumber\\
%&\geq&\Iabs(S:A)+\Iabs(S:B)-I(A:B)\nonumber\\
%&\geq&I(X:A)+I(Z:B)-I(A:B).
%\end{eqnarray}
\begin{eqnarray}
I(S:AB)&=&I(S:A)+I(AS:B)-I(A:B)\nonumber\\
&\geq&I(S:A)+I(S:B)-I(A:B)\nonumber\\
&\geq&I(X:A)+I(Z:B)-I(A:B).
\end{eqnarray}
In this way we proved \eqref{eq:principle_conc_app}.

%{\color{red} zostawic wyrazenie z $I(X_1, B_2, B_3 : B_1 )$ }

%%%%%%%%%%%%%%%%%%%%%%%%%%%%%%%%%%%%%%%%%%%%%%%%%%%%%%%%%%%%%%%%%%%%%%%%%%%
% ICP z wieloma obserwablami
%%%%%%%%%%%%%%%%%%%%%%%%%%%%%%%%%%%%%%%%%%%%%%%%%%%%%%%%%%%%%%%%%%%%%%%%%%%
\section{Information Content Principle for multiple observables \label{inf_principle_general}}
Now we prove \ICP{} for the setup with $n$-observables $\{X_i\}$.
For the classically correlated state (cf.~\eqref{eq:rghSAB})
\begin{equation}
\rho^{SA_1\ldots A_n}=
\sum_{i_1,\ldots,i_n}p_{i_1,\ldots,i_n}\rho^{S}_{i_1,\ldots,i_n}\otimes
\sigma^{A_1}_{i_1}\otimes\ldots\otimes\sigma^{A_n}_{i_n},
\label{eq:rghSAi}
\end{equation}
where $\rho_S$ represents system $\mathcal{S}$ belonging to considered
theory and $\{\sigma^{A_j}_{i_j}\}$ denote classical registers, we show that holds
\begin{equation}
\sum_i I(X_i:A_i)-I(A_1:\ldots:A_n) \leq \log_2 d\label{eq:principle_conc_multi}.
\end{equation}
$d$ is observed system dimension (cf.~(3)) %\eqref{eq:principle_conc}) 
and $I(A_1:\ldots:A_n)=\sum_i H(A_i) - H(A_1,\ldots,H_n)$ is multivariable mutual information.
Upper bound $I(S:A_1,\ldots,A_n)\leq\log_2 d$ comes in exactly the same way as in~\eqref{icp_up} hence we omit this part of proof and focus on LHS of~\eqref{eq:principle_conc_multi}.

We start using chain rule to write:
%\begin{eqnarray}
%\Iabs(S:A_1,\ldots,A_n)&=&\Iabs(S:A_1)+\Iabs(S:A_2|A_1)\nonumber\\
%&&+\Iabs(S:A_3|A_1,A_2)+\ldots+\nonumber\\
%&&\Iabs(S:A_n|A_1,\ldots,A_{n-1}).
%\end{eqnarray}
\begin{eqnarray}
I(S:A_1,\ldots,A_n)&=&I(S:A_1)+I(S:A_2|A_1)\nonumber\\
&&+I(S:A_3|A_1,A_2)+\ldots+\nonumber\\
&&I(S:A_n|A_1,\ldots,A_{n-1}).
\end{eqnarray}
We use chain rule once again to express express conditional mutual information in the form:
%\begin{eqnarray}
%\Iabs(S:A_2|A_1) &=& \Iabs(A_1,S:A_2)\nonumber\\
%&& - I(A_1:A_2)\nonumber\\
%\Iabs(S:A_3|A_1,A_2) &=& \Iabs(A_1,A_2,S:A_3) \nonumber\\
%&&- I(A_1,A_2:A_3)\nonumber\\
%&\ldots&\nonumber\\
%\Iabs(S:A_n|A_1,\ldots,A_{n-1}) &=& \Iabs(A_1,\ldots,A_{n-1},S:A_n) \nonumber\\
%&&- I(A_1,\ldots,A_{n-1}:A_n)\nonumber.
%\end{eqnarray}
\begin{eqnarray}
I(S:A_2|A_1) &=& I(A_1,S:A_2)\nonumber\\
&& - I(A_1:A_2)\nonumber\\
I(S:A_3|A_1,A_2) &=& I(A_1,A_2,S:A_3) \nonumber\\
&&- I(A_1,A_2:A_3)\nonumber\\
&\ldots&\nonumber\\
I(S:A_n|A_1,\ldots,A_{n-1}) &=& I(A_1,\ldots,A_{n-1},S:A_n) \nonumber\\
&&- I(A_1,\ldots,A_{n-1}:A_n)\nonumber.
\end{eqnarray}
Combining these together with strong subadditivity we get:
%\begin{eqnarray}
%\Iabs(S:A_1,\ldots,A_n)&\geq&\Iabs(S:A_1)+\ldots+\Iabs(S:A_n)\nonumber\\
%&&-I(A_1:A_2)-\ldots\nonumber\\
%&&-I(A_1,\ldots,A_{n-1}:A_n).\label{iabs_multi_1}
%\end{eqnarray}
\begin{eqnarray}
I(S:A_1,\ldots,A_n)&\geq&I(S:A_1)+\ldots+I(S:A_n)\nonumber\\
&&-I(A_1:A_2)-\ldots\nonumber\\
&&-I(A_1,\ldots,A_{n-1}:A_n).\label{iabs_multi_1}
\end{eqnarray}
From the classical mutual information properties ($I(A:B)=H(A)+H(B)-H(A,B)$),
it is easy to see that $I(A_1:A_2)+\ldots+I(A_1,\ldots,A_{n-1}:A_n)=I(A_1:\ldots:A_n)$ holds.
Putting that to~\eqref{iabs_multi_1} and applying information processing inequality for measurements, we finally get:
%\begin{equation}
%\Iabs(S:A_1,\ldots,A_n)\geq\sum_i I(X_i:A_i)-I(A_1:\ldots:A_n).\nonumber
%\end{equation}
\begin{equation}
I(S:A_1,\ldots,A_n)\geq\sum_i I(X_i:A_i)-I(A_1:\ldots:A_n).\nonumber
\end{equation}
That finishes the proof.

\section{Entropy in \GPT{} \label{sec:entropy_props}}
%{\color{red} pokazac perfect decodable odpowiada wymiarowi... i dyskusja entropii}
In this section we would like to focus on the properties (i)-(iv) of entropy which was used in the proofs presented in Appendix~\ref{supplement material_ICP_proof} and Appendix~\ref{inf_principle_general}.

We may define some general notion of entropy $\mathcal{H}$ which measure our uncertainty about the system $S$ which belong to \GPT{}.
The natural assumption is that $\mathcal{H}$ should reduce to classical or quantum entropy if we restrict to these theories.

%We discuss some assumption which leads to these properties on the ground of
%\GPT{}.
%The discussion should provide some insight into assumptions which are broken in .
Moreover, as it was pointed out in \cite{GPT_Entropy_conditions}, properties (iii) and (iv) follow from resonable assumption that local transformation can destroy but not create correlations. This assumption express in the formal way as:
\begin{equation}
\Delta H(A B) \geq \Delta H(A),
\end{equation}
where the transformation is performed on the system $A$.
We expect that the theory provides at least transformations like system preparation, measurement and discarding.

Property (ii) refers to the procedure of system preparation where we
randomly choose one of the several possible state of the system.
The knowledge of the way how the system was prepared should reduce our uncertainty.

To motivate (i), first we would like bring to attention that the general entropy $\mathcal{H}$ is often linked with the minimal output uncertainty on the distinguished subset of measurements $\mathcal{M}_F$:
\begin{equation}
\mathcal{H}(S)=\inf_{M\in\mathcal{M}_F} H(M(S)),\label{entropy_output}
\end{equation}
where $M$ is measurement on the system $S$.
This distinguished subset $\mathcal{M}_F$ consist of maximally informative, i.e. fine-grained measurements\cite{GPT_entropies,GPT_entropies2}.
In the analogy to quantum mechanics we may think of them as a set of rank-1 POVMs.

In the quantum case special role is played by the projective measurements.
The von Neuman entropy is the output entropy for the measurement which consists of projectors on the eigenvectors of the state. In $\GPT{}$ we call a measurement a projective measurement if for every outcome $e$ there is a state $\omega$ that the probability  of the outcome $e$ on the state $\omega$ is $e(\omega)=1$. We observe that non-projective measurements contains some intrisic noise, i.e. some outcomes cannot be obtained with probability one. On the other hand, information encoded in
states $\{\omega_j\}$ can be perfectly retrieved since $e_i(\omega_j)=\delta_{i,j}$.

For that reasons, we assume that entropy should refer to the uncertainty of the outcome of fine-grained projective measurements and in this way it should be bounded by the number of bits $d$ that may be encoded in the system in perfectly decodable way.

Interestingly, with some additional assumptions on the post measurement state, if $\mathcal{H}(S)$ attains its value for projective measurement, then it has operational interpretation in terms of information compression\cite{GPT_entropies2}.

%%%%%%%%%%%%%%%%%%%%%%%%%%%%%%%%%%%%%%%%%%%%%%%%%%%%%%%%%%%%%%%%%%%%%%%%%%%
% ICP z wieloma obserwablami
%%%%%%%%%%%%%%%%%%%%%%%%%%%%%%%%%%%%%%%%%%%%%%%%%%%%%%%%%%%%%%%%%%%%%%%%%%%

\section{Violations of \ICP{} in general probabilistic theories - details\label{sec:UPandRAC_app}}
In this section we provide technical details which support discussion on
violations of \ICP{} in general probabilistic theories which was presented in main part of the paper.
We will base on the fact that normalized states of $p$-GNSTs and polygon theories are real vectors $\omega\in\mathbb{R}^2$. Maximal number of perfectly distinguishable states thus satisfies
$d\leq 3$. Equality holds if and only if the states space is a simplex \cite{GPT-fine_obs_MullerDV}.
Therefore any non-classical theory has $d\leq 2$ so that the information content of the system satisfies $\mathcal{I}_C\leq 1$.

%{\color{red} dopisa� ogulnie o zwi�zku z zasadami nieoczaczono�ci - kodowanie w niezale�ne A i B, maksymalna entropia na stanie zredukowanym}

\subsection{Violation uncertainty relation for anti-commuting observables\label{sec:pGNSTdetails}}
We consider the $p$-GNST with two dichotomic observables $X$ and $Z$. Admissible states fulfill uncertainty relation:
\begin{equation}
(s_x)^p+(s_z)^p \leq 1,\label{eq:ptheory_ucrel}
\end{equation}
where $p\in[2,\infty]$ is parameter of the theory and $s_x=p(a = +|x = X)_\psi-p(a = -|x = X)_\psi$ is the mean value of observable $X$ measured on the system in state $\psi$ (analogically for $s_z$ and $Z$). It is straightforward to see that~\eqref{eq:ptheory_ucrel} is an uncertainty relation since it bounds the probability that the state has well defined outcome of each observable.
p-GNST is a simplified version of the model discussed in \cite{SW_relaxed_uncertainty} since we only deal with the case of $2$ observables available in the elementary system. However our results may be easily generalized to the case of $3$ observables $X,Y,Z$.

The set of admissible states for $p=2$ correspond to the set of states from the great circle of the Bloch ball (in case of $3$ observables, the set of admissible states becomes full Bloch ball and relation of type~\eqref{eq:ptheory_ucrel} define the state space of single qbit). On the other hand, for $p\rightarrow\infty$ we approach to state space of sbit.
Therefore increase of $p$ leads to relaxation of the uncertainty relation.

Now we show, that each theory with $p>2$ violates \ICP{}. For that purpose, (i) we show that there exist a state $\psi_{++}$ with entropic uncertainty small enough (i.e. $H(X)_{\psi_{++}} + H(Z)_{\psi_{++}} < 1$); (ii) then by symmetry of the state space we construct state $\rho^{SAB}$ which we use to prove violation.

Let us parameterize by $s_x$ states $\psi$, that saturate~\eqref{eq:ptheory_ucrel}. For simplicity we assume that $s_z>0$. Due to~\eqref{eq:ptheory_ucrel}, we have $s_z=\sqrt[p]{1-s_x^p}$.
For $s_x=1$, the outcome of the observable $X$ is certain but we have no knowledge on the outcome of observable $Z$. As $s_x$ decrease, the knowledge on the outcome of $Z$ increase by the cost of certainty of the outcome of $X$. Rate of this exchange depends on the uncertainty relation and interestingly for $p>2$, some states near to $s_x=1$ have entropic uncertainty smaller than in the quantum case. Precisely, we show
%in SEC.~\ref{small_entrop_uncert}
that there exist $\delta_x$ that any state with $1-\delta_x < s_x < 1$, fulfill:
\begin{equation}
H(X)_\psi + H(Z)_\psi < 1.
\label{eq:smallentropy}
\end{equation}

For parametrization of state $\psi$ by $s_x$, entropies of measurements take a form
$H(X)_{\psi} = H(\frac{s_x+1}{2}), H(Z)_{\psi}=H(\frac{s_z+1}{2})$ and
are bounded in the following way:
$H(\frac{s_x+1}{2})\leq (\frac{1-s_x}{2})^{1+\epsilon}$,
for $\epsilon>0$ and $\frac{1-s_x}{2} < \delta_\epsilon$ and
$H(\frac{s_z+1}{2})\leq 1 - (\frac{s_z}{2})^2$.
This allows us to rewrite condition~\eqref{eq:smallentropy} as:
\begin{equation}
\left(\frac{1-s_x}{2}\right)^{1+\epsilon} < \frac{1}{4} \left(1-s_x^p\right)^{2/p}.
\label{eq:smallentropyapprox}
\end{equation}
Now let us observe that:
\begin{equation}
\left(\frac{1-s_x}{2}\right)^{1+\epsilon}\Big|_{s_x=1}=\frac{1}{4} \left(1-s_x^p\right)^{2/p}\Big|_{s_x=1}=0,
\end{equation}
and for $(1+\epsilon)p>2$:
\begin{equation}
\lim_{s_x\rightarrow 1} \frac{(1-s_x^p)^{2/p}}{(1-s_x)^{1+\epsilon}} =\infty.
\end{equation}
It means that LHS of~\eqref{eq:smallentropyapprox} converge to $0$ faster than RHS as $s_x\rightarrow 1$. Since both sides of~\eqref{eq:smallentropyapprox} are positive, it implies that~\eqref{eq:smallentropyapprox} holds for $1-\delta_x < s_x < 1$ with $\delta_x$ small enough.

Since we have shown that states with desired property exists, we can
take any state $\psi_{++}$ that $H(X)+H(Z) = \tilde{H}<1$. The state is described by
$(\tilde{s}_x,\tilde{s}_z)$. By the symmetry of~\eqref{eq:ptheory_ucrel} and~\eqref{eq:smallentropy}, we know that states $\psi_{+-},\psi_{-+},\psi_{--}$ obtained from $\psi_{++}$ by negation of proper parameter are also admissible and have the same entropic uncertainty $\tilde{H}$.
This allow us to construct the state:
\begin{equation}
\rho^{SAB}=\frac{1}{4}\sum_{i,j\in\{-,+\}}\psi_{ij}^S\otimes i^A\otimes j ^B.
\label{p-state}
\end{equation}
It is easy to observe that outcome of $X$ and $Z$ for reduced state $\frac{1}{4}\sum_{i,j\in\{-,+\}}\psi_{ij}^S$ is completely random.
Hence we may write:
\begin{eqnarray}
&&I(X:A)+I(Z:B)\nonumber\\
&&=H(X)+H(Z)\nonumber\\
&&-\!\!\sum_{i,j\in\{-,+\}}\!\!\frac{1}{4}(H(X)_{\psi_{i,j}}+H(Z)_{\psi_{i,j}})\nonumber\\
&&=2-\tilde{H}\nonumber\\
&&>1.
\end{eqnarray}
Since, in addition, from (\ref{p-state}) we have  $I(A:B)=0$  (\ref{p-state}) the above shows the expected violation and finishes the proof.

\subsection{Polygon theories\label{sec:polygondetails}}
\label{polygon}
Polygon theories (parameterized by $n$) was developed in \cite{polytopic_models} to study connection between the strength of non-local correlations and the structure of the state spaces of individual systems. They may be viewed as a progressive relaxation of superposition principle (c.f. relaxation of uncertainty relation in p-GNSTs) moving from quantum case $n\rightarrow\infty$ to sbit ($n=4$) and classical trit ($n=3$). Relaxation of superposition principle means that more restriction are putted on the way the states can be superposed.

%Fine grinned measurements consists of effect $e_i$ and its "complementary" effect $\bar{e}_i=u-e_i$ with exception of $n=3$ where it has form $\{e_1,e_2,e_3\}$ \cite{GPT_conic,no_info_gain}. We do not consider this case.

The proof of violation of \ICP{} by unphysical (i.e. with $n>3$ and $n<\infty$) polygon theories is quite technical and base mostly on construction of state $\rho^{SAB}$ with proper measurement entropies. We start with short description of polygon theories mainly following \cite{polytopic_models}. For more details see original paper.

State space $\Omega$ of a single system in polygon theory is a regular polygon with $n$ vertices. For fixed $n$, $\Omega$ is represented as a convex hull of $n$ pure states $\{\omega_i\}_{i=1}^n$:
\begin{equation}
	\omega_i=	
	\begin{pmatrix}
	r_n\cos\left(\frac{2 i \pi}{n}\right)\\
	r_n\sin\left(\frac{2 i \pi}{n}\right)\\
	1
	\end{pmatrix}
	\in \mathbb{R}^3,
\end{equation}
where $r_n=1/\sqrt{\cos(\pi/n)}$.

The set of effects is the convex hull of the unit effect, zero effect and the extreme effects. The unit effect has form:
\begin{equation}
	u=	
	\begin{pmatrix}
	0\\
	0\\
	1
	\end{pmatrix}.	
\end{equation}
Extreme effects for even $n$ are given by:
\begin{equation}
	e_i=	\frac{1}{2}
	\begin{pmatrix}
	r_n\cos\left(\frac{(2i-1)\pi}{n}\right)\\
	r_n\sin\left(\frac{(2i-1)\pi}{n}\right)\\
	1
	\end{pmatrix}
	\in \mathbb{R}^3,
\end{equation}
and for odd $n$ in slightly different form:
\begin{equation}
	e_i=	\frac{1}{1+r_n^2}
	\begin{pmatrix}
	r_n\cos\left(\frac{2i\pi}{n}\right)\\
	r_n\sin\left(\frac{2i\pi}{n}\right)\\
	1
	\end{pmatrix}, e'_i = u - e_i
	\in \mathbb{R}^3.
\end{equation}
$e(\omega)=e\cdot\omega$ is the Euclidean inner product of the vectors representing the effect and the state.

Now we are in position to construct states which violate Information Content principle in the polygon theories.
We will consider separately the case of even and odd $n$.

For even $n$ we use the state
\begin{eqnarray}
\rho^{SAB}&=&\frac{1}{4}\left(\omega_2^S\otimes\sigma_0^A\otimes\sigma_0^B \right.\nonumber\\
&&+ \omega_1^S\otimes\sigma_0^A\otimes\sigma_1^B\nonumber\\
&&+ \omega_{n/2+1}^S\otimes\sigma_1^A\otimes\sigma_0^B\nonumber\\
&&+ \left.\omega_{n/2+2}^S\otimes\sigma_1^A\otimes\sigma_1^B\right)
\end{eqnarray} along with
measurement $X$ and $Z$ given by effects $\{e_2, u-e_2\}$ and  $\{e_{\left\lfloor n/4 \right\rfloor+2}, u - e_{\left\lfloor n/4 \right\rfloor+2}\}$ respectively. It is easy to see that $I(A:B)=0$ since
each combination $\sigma_i^A\otimes\sigma_j^B$ occurs with the same probability $1/4$.
To calculate $I(X:A)$ and $I(Z:B)$ we need conditional entropy of measurement outcome which may be obtained from probability of given effects for particular state (i.e. $e_j(\omega_i)$). For $I(X:A)$ the probabilities are $e_2(\omega_1)=e_2(\omega_2)=1$ and $e_2(\omega_{n/2+1})=e_2(\omega_{n/2+2})=0$, hence $I(X:A)=1$. For $I(Z:B)$, straight forward calculations lead to:
\begin{eqnarray}
p(Z=0|B=0)&=&\nonumber\\
p(Z=1|B=1)&=&\frac{1}{2}\left(1+\sin\left(\frac{2\pi\left\lfloor\frac{n}{4}\right\rfloor}{n}\right)\tan\left(\frac{\pi}{n}\right)\right)\nonumber.
\end{eqnarray}
It shows that $I(Z:B)>0$, hence the violation of the Information Content principle was proved.

For odd $n$ we use the state
\begin{eqnarray}
\rho^{SAB}&=&\frac{1}{4}\left(\omega_1^S\otimes\sigma_0^A\otimes\sigma_0^B \right.\nonumber\\
&&+ \omega_1^S\otimes\sigma_0^A\otimes\sigma_1^B\nonumber\\
&&+ \omega_{\left\lfloor n/2\right\rfloor+1}^S\otimes\sigma_1^A\otimes\sigma_0^B\nonumber\\
&&+ \left.\omega_{\left\lfloor n/2\right\rfloor+2}^S\otimes\sigma_1^A\otimes\sigma_1^B\right).
\end{eqnarray}
In this case measurement $X$ and $Z$ are given by effects $\{e_1, u-e_1\}$ and  $\{e_{\left\lfloor n/4 \right\rfloor+1}, u - e_{\left\lfloor n/4 \right\rfloor+1}\}$ respectively.
Once again we have that $I(A:B)=0$ and $I(X:A)=1$. Formulas for $p(Z=0|B=0)$ and $p(Z=1|B=1)$ are more complicated:
\begin{eqnarray}
p(Z=0|B=0)&=&\frac{1}{4} \left(
2 \cos\left(\frac{\pi }{n}\right)+
\cos\left(2 \pi\frac{\left\lfloor\frac{n}{4}\right\rfloor}{n}\right)+\right.\\
&&\left.
\cos\left(2 \pi\frac{ \left\lfloor\frac{n}{4}\right\rfloor-
\left\lfloor\frac{n}{2}\right\rfloor}{n}\right)\right) \sec\left(\frac{\pi }{2 n}\right)^2\nonumber\\
p(Z=1|B=1)&=&\frac{1}{4} \left(2-\cos\left(2 \pi\frac{ \left\lfloor\frac{n}{4}\right\rfloor}{n}\right)-\right.\\
&&\left.\cos\left(2 \pi \frac{  \left\lfloor\frac{n}{4}\right\rfloor-\left\lfloor\frac{n}{2}\right\rfloor-1}{n}\right)\right) \sec\left(\frac{\pi }
{2 n}\right)^2\nonumber.
\end{eqnarray}
However we get that $p(Z=0|B=0)>1/2$ and $p(Z=1|B=1)>1/2$ hence $I(Z:B)>0$ that proves violation also in this case.

% therefore we moved it to SEC. \ref{polygon_states} and  here we discuss only the results.

\ICP{} violation in polygon theories is connected, as in the case of $p$-GNST, with uncertainty relations. It is easy to see especially for even $n$. We notice that for $n=4m+2$, where $m$ is integer, non-complementary observables are measured.
In case of odd $n$, role of uncertainty is less obvious because of asymmetry of the state $\rho^{SAB}$.

Correlations obtained in models with odd $n$ do not violate Tsirelson's bound \cite{polytopic_models}. It means that this class of theories cannot be separated from the quantum theory using standard argumentation \cite{inf_causality}.
%Limitation on non-locality is connected with fact that maximally entangled states in these theories are inner product states and lead to correlations which belong to $Q^1$ \cite{polytopic_models}.
Since non-locality is tightly connected with uncertainty relations  it might be interesting to 
apply \ICP{} to explain the impossibility of steering to maximally certain states \cite{uncert_nonloc}.
%Violation of \ICP{} suggest more direct connection of \ICP{} and uncertainty relations.

 Very recent results on the classical information transmission in polygon theories \cite{GPT_holevo_polygon} provide some more insight into this issue.
It turns out that polygon theories with odd $n$ allow for communication of more than $1$-bit per elementary system in Holevo sense (i.e. in asymptotic limit). It means that \eqref{eq:principle_conc_app} is violated even in one observable setup when non-pure measurement is performed. Therefore our result for odd $n$ may be viewed as a simple consequence of the fact, that Holevo like capacity exceeds number of bits which may be encoded in the system in perfect decodable way. This is contrary to what we observe for classical and quantum systems.
For even $n$, Holevo like capacity of the elementary system is $1$-bit.
It emphasizes the advantage of multiobservable approach over Holevo like in discrimination of non-local theories. It is interesting if for odd $n$, information content for multiple observables may exceed Holevo limit.

%{\color{red} Very recent results show that polygon theories with odd $n$ allow for communication of more than $1$-bit per elementary system in Holevo sense (i.e. in asymptotic limit) and it links to non-pure measurements. For even $n$, only $1$-bit of information may be transmitted. In this context,
%replacing single system approach expressed by "perfect decodability" with Holevo like capacity for given theory will relax \ICP{} and make it more consistent with correlation approach and Tsirelson's bound. This will be discussed in other paper.}

%1) wiaze sie z zasadna nieoznaczonosci, tylko ze nie koniecznie komplementarnych obserwabli
%2) co ciekawe wykrywa takze dla nieparzystych, steering w stany o nieoznaczonosci

At the end of this section we use polygon theories to compare \ICP{} with criterion based on the
mismatch between measurement dimension and information dimension\cite{Brunner_dimensions}.
Here measurement dimension denotes the number of perfectly decodable states and
information dimension the number of pairwise perfectly distinguishable states.
For polygon theories with $n\in\{4,\ldots,13\}$, mismatch between
measurement dimension and information dimension take place only for $n\in\{4,6\}$. Using this approach only two cases may be ruled out while \ICP{} rules out all of them. However, it cannot be excluded, that the mismatch criterion will rule out these theories, if we consider composite systems, with appropriate choice of composition rules.

\subsection{Composite systems}
At this point we go back to p-GNSTs. We will consider p-GNSTs in their original formulation
from~\cite{SW_relaxed_uncertainty}, i.e. where $3$ dichotomic and anticommuting observables may be measured on elementary system.

We show that \ICP{} is able to exclude non-local \GPT{} even if elementary system state space is a Bloch sphere. Namely, we show that for p-GNST with $p=2$, big enough multipartite system violate \ICP{}. For this purpose we take advantage of super-strong \RAC{} present in that theories.

%We also demonstrate, that despite \ICP{} is formulated in terms of elementary system, it naturaly generalize to composite systems.
This will also demonstrate that \ICP{} can be applied to composite systems.

As it was shown in~\cite{SW_relaxed_uncertainty}, p-GNST with $p=2$ allows for encoding $3^n$ bits in $n$-gbit state with single bit recovery probability equal $p_{rec}=\frac{1}{2}+\frac{1}{2\sqrt{2n+1}}$.
Since each bit is decoded by different observable, and the bits are distributed uniformly and independently we obtain:
\begin{equation}
\sum_{i}I_{M_i}-I_R = \sum_{i=1}^{3^n} I(O_i:A_i)=3^n \left( 1 - H(p_{rec}) \right),
\end{equation}
where $i$ denotes bit $A_i$ which is decoded by observable $O_i$ and $H$ is classical entropy.
On the other hand, for $n$-gbit system, maximal number of perfectly decodable states is bounded by $d\leq (3+1)^n$ (cf.~\eqref{eq:state_space_dimm} and~\eqref{eq:dimbound}). Putting this together we get that $5$-gbit system violate \ICP{} (i.e $\sum_{i}I_{M_i}-I_R = 16.18 > 10=\log_2 4^5$).
This result relates to~\cite{local_qm_and_ns,unif_framework} where it was shown that locally
quantum state space with no-signaling conditions implies fully quantum state space for bipartite systems, however the situation changes dramatically in multipartite scenarios.

The difference $\sum_{i}I_{M_i}-I_R$ depends strongly on the state space of composite system. As we have seen, the uncertainty relations for anticommuting observables do not restrict state space strongly enough to ensure that \ICP{} is satisfied.
Therefore it is interesting to ask what other geometrical constraints (e.g. other uncertainty relations, consistency constraints (cf.~\cite{SW_relaxed_uncertainty}), local orthogonality) have to be added to the theory to conform with~\ICP{}. In the opposite direction, one may ask how \ICP{} limits strength of non-local correlations achievable in \GPT{} for system whose elementary subsystems obey quantum mechanics.

\section{ {\it\ ICP} versus existing axioms and principles\label{sec:ICP_versus}}

It should be noted here, that postulating \ICP{}  we do not search for "physical" justification for inequality (4), 
%\eqref{eq:principle_conc_multiple} 
but rather in spirit of information theoretical principles such as Information Causality \cite{inf_causality} (IC) or local orthogonality \cite{local_orthogonality} we aim for understanding  the physical reality by means of information approach. In this  context it is natural to ask how ICP is related to the existing principles and axioms. As we mentioned in the introduction there are,  in substance, two paradigms within of we try to understand the peculiar role of quantum mechanics in the set of possible theories of physical world. The first one is to derive quantum mechanics from more intuitive axioms. The other is to pose  a single principle which is more complicated  than simple axioms but has a chance to, at least, rule out  many theories with different predictions than quantum ones. For sake of clarity we will reefer further these two different paradigms as to "axioms" and "principles". Note that these two paradigms are not   comparable. Clearly any principle can be derived from axioms as they reproduce quantum mechanics. However there is never simple connection between axioms and principles. For instance the IC is now know not to be capable to reproduce quantum mechanics (i.e. to be "worse" than axioms) but clearly does not mean that it is indeed less important. 

      Since the crux of our constrained uncertainty  principle is its information theoretic flavor, and usage of  strong subadditivity that holds for both quantum and classical world it cannot be a simple consequence of axioms such as e.g. \cite{qm_deriv_mm,qm_deriv_chiribella}. But how it is related to  existing principles? There is a basic difference between them.  Our principle applies to single system, while all the existing principles involve correlations between subsystems, hence cannot be applied to a single system.

%
%- consistency
%- interpretacja jako XOR game (??) - wystepuja silne korelacje wieloczastkowe
%- dlaczego dopiero dla 4
%- pokazalismy ze jest takze ze mimo iz sformulowanie jest single partite to mozna je wykozystac
%w schemacie multipartite
%- nieoznaczonosc wieloczastkowa (??) nie jest wystarczajaco cisasna ??
%- dopisac uklady zlozone w GPT

%{0.767977, 1 , 2},  <= tak jak w kwantach
%{1.34559, 2 , 4},  <= ci�gle spe�nia nasza zasade (ale to wynika z tego ze uklad 2-czastkowy lokalnie %kwantowy tez jest kwantowy)
%{2.85269, 3, 6 , 4.7549}, <= to byc moze wynika z tego ze w p-GNST mamy pewne wiezy wieloczastkowe
%{6.61804, 4 , 8, 6.3398},  <= lamanie naszej zasady
%{16.1859, 5, 10}

%%%%%%%%%%%%%%%%%%%%%%%%%%%%%%%%%%%%%%%%%%%%%%%%%%%%%%%%%%%%%%%%%%%%%%%%%%%
% czesci techniczne dowodzenia lamania ICP w roznych teoriac
%%%%%%%%%%%%%%%%%%%%%%%%%%%%%%%%%%%%%%%%%%%%%%%%%%%%%%%%%%%%%%%%%%%%%%%%%%%
%\section{Existence of states with small entropic uncertainty \label{small_entrop_uncert}}
%This section contain technical part where we proof that for $p>2$ there exist state $\psi$ in p-GNST that
%\begin{equation}
%H(X)_\psi + H(Z)_\psi < 1.
%\label{eq:smallentropy2}
%\end{equation}

%\section{States violating \ICP{} in polygon theories \label{polygon_states}}
\clearpage

%\end{supplemental material}
\end{appendix}

\end{document}